\documentclass{aastex63}

\usepackage{gensymb}
\usepackage{}

\submitjournal{ApJL}

\shorttitle{Phaethon - Mass shedding activities }
\shortauthors{Nakano \& Hirabayashi.}

\graphicspath{{./}{figures/}}

\begin{document}

\title{Mass shedding activities of Asteroid (3200) Phaethon enhanced by its rotation}

\correspondingauthor{Ryota Nakano}
\email{rzn0040@auburn.edu}

\author{Ryota Nakano}
\affiliation{Department of Aerospace Engineering \\
Auburn University, 211 Davis Hall\\
Auburn, AL 36849, USA}

\author{Masatoshi Hirabayashi}
\affiliation{Department of Aerospace Engineering \\
Auburn University, 211 Davis Hall\\
Auburn, AL 36849, USA}

\begin{abstract}
Asteroid (3200) Phaethon, a B-type asteroid, has been active during its perihelion passages. This
asteroid is considered to be a source of the Geminid meteor stream. It is reported that this asteroid is
spinning at a rotation period of $3.60 \ hr$ and has a top shape (an oblate body with an equatorial ridge)
with a mean equatorial diameter of $6.25 \ km$. Here, we report that Phaethon's rotation state may be close to
or above its critical rotation period when the bulk density is $0.5 \ - \ 1.5 \ {g/cm^3}$ (a typical bulk
density of a B-type asteroid). We found that in this condition, the structure of Phaethon is sensitive to
failure unless the cohesive strength is ${\sim}50 \ Pa \ - \ {\sim}260 \ Pa$. This result implies that if
there are some surface processes driven by, for example, thermal waves, large-scaled deformation may happen
and cause mass shedding. From this interpretation, we propose the processes that produced the Geminid
meteor stream in the past and dust tails recently. Phaethon initially rotated at a spin period shorter than
the current period. The magnitude of structural deformation at this stage was higher than the present spin
condition, and a large mass shedding event, i.e., the Geminid meteor stream, occurred. After this deformation
process, the body became more oblate, and its spin slowed down. At this point, while the spin was high enough
for the body to have mass shedding events, the magnitude of these events became small.
\end{abstract}


\section{Introduction} \label{sec:intro}

Asteroid (3200) Phaethon may be a source of the Geminid meteor stream \citep{Whipple1983, Gustafson1989, Williams1993}
and has been observed for a long time \citep{Cochran1984, Chamberlin1996, Hsieh2005, Wiegert2008}. In 2009
\citep{Jewitt2010} and 2012 \citep{Li2013}, dust tails from Phaethon were observed during the perihelion passage,
revealing that this asteroid is indeed an active asteroid. However, the activities detected near the perihelion left a
mystery of the production of the Geminid meteor stream. The dust mass inferred from the observations is
${\sim}3\times10^5 \ kg$ \citep{Li2013}, which is much smaller than the estimated mass of the Geminid meteor stream,
${10}^{12}-{10}^{13} \ kg$ \citep{Hughes1989, Jenniskens1994}. Also, the estimated average mass-loss rate of 
$3\ {kg}/s$ is too small to replenish the Geminid meteor stream mass \citep{Jewitt2013}, if the dynamical lifetime of
the Geminid meteor stream is ${\sim}10^3 \ yr$ \citep{Gustafson1989, Ryabova2007}. While most of the proposed
mechanisms were found incapable of producing dust tails, thermal fracture and cracking due to dehydration in surface
materials might be reasonable processes to generate dust tails \citep{Jewitt2010, Jewitt2012, Li2013}. 

Radar observations during the 2017 apparition revealed Phaethon’s shape. \citet{Taylor2019} reported that Phaethon might
be an oblate shape with an equatorial ridge, or the so-called top shape. The equivalent diameter of this asteroid was
estimated to be $6 \ km$ in diameter. The examples of top-shaped asteroids are OSIRIS-REx’s target (101955) Bennu
\citep{Lauretta2019} and Hayabusa2’s target (162173) Ryugu \citep{Watanabe2019, Sugita2019, Kitazato2019}. Phaethon is
currently spinning at a rotation period of $3.60 \ hr$ \citep{Taylor2019}. The radar albedo is reported to be the lowest
among the cataloged near-Earth asteroids \citep{Taylor2019}, implying that its spectral type is consistent with a B-type
\citep{Taylor2019}, and thus the bulk density may be as low as ${\sim}1.0 \ g/{cm}^{3}$
\citep{Scheeres2019, Watanabe2019}.

We hypothesize that the mass shedding activities of Phaethon may have been enhanced by fast rotation, given recent work
arguing that the equatorial ridges of top-shaped asteroids were evolved by rotationally driven reshaping
\citep{Walsh2008, Walsh2012}. We propose that structural failure on the surface and/or inside the body triggered
by fast rotation plays an important role in mass shedding mechanism. This study provides better
insights into the physical activities of Phaethon to support DESTINY+, a planned flyby mission concept led by
the Japan Aerospace Exploration Agency \citep{Arai2018}. 

\section{Semi-analytical model for structural failure in a top-shaped body} \label{sec:Semi-analytical model}

Phaethon was reported to be a top-shaped body. Recent work has shown that the global failure condition
in uniformly rotating top shapes can be roughly determined by assuming that they are triaxial ellipsoids
\citep{Hirabayashi2015}. For Phaethon, the shape is not
well known at present while an effort in analyzing it from radar observation data is being made (Taylor,
personal communication). Therefore, a simplified model that uses a triaxial ellipsoid can still reasonably
provide structural failure in this asteroid. We note that the heterogeneity in structural failure may be critical once the detailed shape of this asteroid is considered \citep{Hirabayashi2019}.

We consider that Phaethon has a mean equatorial diameter of $6.25 \ km$ and an oblateness (the ratio of
semi-minor axis to semi-major axis) of $0.889$, which is the same as Bennu's \citep{Barnouin2019}. We note that
\citet{Taylor2019} did not specify the oblateness of Phaethon but implied that it would be similar to that of Bennu. 
For the oblateness, we do not account for the semi-intermediate axis to simplify the discussion. We assume that
Phaethon’s structure is uniform because the internal condition is unknown. We consider three types of the bulk density,
$0.5 \ g/{cm}^3$, $1.0 \ g/{cm}^3$, and $1.5 \ g/{cm}^3$, the averaged of which is consistent with that of a B-type
asteroid \citep{Scheeres2019}. Later, we denote the oblateness, the bulk density, and the gravitational constant as
$\epsilon$, $\rho$, and $G$, respectively. Phaethon is assumed to be rotating along the maximum principal axis. We
define a three-dimensional Cartesian coordinate system such that the $z$ axis is lined up along the rotation axis, and the
$x$ and $y$ axes are along the maximum and intermediate moment of inertia axes, respectively, on the equatorial plane.
Using these definitions, we compute how Phaethon experiences structural failure at a give rotation period, $P$.

To determine the failure condition of Phaethon, we extend the technique by \citet{Hirabayashi2015}, who only considered
a sphere. We analyze when the stress field in a given element reaches its yield condition. In this model, the material
behavior is assumed to be elastic-perfectly plastic, where a plastic flow begins once the stress reaches its yield
condition without material hardening and softening. To describe such a material behavior, we use the following material
properties: Poisson’s ratio, $\nu$; Young’s modulus, $E$; the angle of friction, $\phi$; and the cohesive strength, $Y$.
It is worth noting that the evolution of plastic deformation is a function of loading paths. However, our purpose is not
to track plastic deformation but to determine what element would first have its plastic state in a quasi-static condition
where deformation is small enough that the time-variation is negligible \citep{Hirabayashi2015}.

Although we follow the terminologies of the cohesive strength by \citet{Hirabayashi2015}, we reintroduce them here to 
facilitate the following discussion. We compute the minimum cohesive strength that can prevent structural failure of a
given element in an asteroid rotating at $P$. We call this strength ‘critical cohesive strength’ and denote it as $Y^*$.
On the other hand, we use ‘actual cohesive strength’ as an assumed strength that an asteroid may have. We denote this as
$Y$. Also, the critical rotation period $P_c$ is the rotation period at which a small particle on the equatorial surface
of Phaethon gains the centrifugal acceleration larger than the gravitational acceleration and is lifted off from the
surface. 

\subsection{Stress field computation} \label{Stress field computation}

Similar to \citet{Hirabayashi2015a} and \citet{Hirabayashi2015}, we apply a technique by \citet{Dobrovolskis1982} and
\citet{Holsapple2001} to provide an elastic stress in a triaxial ellipsoid that is uniformly spinning at rotation period
of $P$. Here, while noting that the details are found in \citet{Dobrovolskis1982}, we briefly introduce the formulation
process. The displacement $u$ in Cartesian coordinates can be expressed in terms of twelve unknown constants $A$ through
$L$:

\begin{equation}
    u_x = x \left[ A + B\frac{x^2}{a^2} + C\frac{y^2}{b^2} + D\frac{z^2}{c^2} \right],
\end{equation}
\begin{equation}
    u_y = y \left[ E + F\frac{x^2}{a^2} + G\frac{y^2}{b^2} + H\frac{z^2}{c^2} \right],
\end{equation}
\begin{equation}
    u_z = z \left[ I + J\frac{x^2}{a^2} + K\frac{y^2}{b^2} + L\frac{z^2}{c^2} \right].
\end{equation}

\noindent where $a$, $b$, and $c$ $(a \geq b \geq c)$ are the principle semi-axes of the triaxial ellipsoid.
The strain is obtained by:

\begin{equation}
    \epsilon_{ij} = \frac{1}{2} \left( \frac{\partial u_i}{\partial x_j} + \frac{\partial u_j}{\partial x_i} \right).
\end{equation}

\noindent Applying Hooke's law, the stress tensor is also expressed in terms of $A$ through $L$:

\begin{equation}
    \sigma_{ij}= \lambda \epsilon_{kk} \delta_{ij} + 2\mu \epsilon_{ij},
\end{equation}

\noindent where $\epsilon_{kk} = \epsilon_{xx} + \epsilon_{yy} + \epsilon_{zz}$, and $\delta_{ij}$ is the
Kronecker delta. $\lambda$ and $\mu$ are the Lame's constants obtained from

\begin{equation}
    \lambda = \frac{E \nu }{(1+\nu)(1-2\nu)},
\end{equation}
\begin{equation}
    \mu = \frac{E}{2(1+\nu)}.
\end{equation}

To determine the twelve unknown constants, we must impose twelve linearly independent relations. The stresses
$\sigma_{ij}$ in a body in equilibrium under body forces must satisfy the stress equilibrium equations:

\begin{equation}
    \frac{\partial}{\partial x_j} \sigma_{ji} = -\rho b_i,
\end{equation}

\noindent where $b_i$ is the body force, which is driven by gravitational and rotational effects in our
problem. Equation (8) provides three constraints out of the twelve required relations. Remaining nine
relations are then imposed from the traction free boundary condition:

\begin{equation}
    \sigma_{ij} n_j = 0,
\end{equation}

\noindent where $n_j$ is the unit normal to the surface and given by:

\begin{equation}
    n_x = \frac{x}{a^2 w},
\end{equation}

\begin{equation}
    n_y = \frac{y}{b^2 w},
\end{equation}

\begin{equation}
    n_z = \frac{z}{c^2 w},
\end{equation}

\noindent where

\begin{equation}
    w = \left( \frac{x^2}{a^4} + \frac{y^2}{b^4} + \frac{z^2}{c^4}\right)^{1/2}
\end{equation}

\noindent The twelve unknown constants are, hence, provided by twelve linearly independent relations.

\subsection{Structural failure condition} \label{subsec:strucutral failure condition}
Once the stress field is obtained from the previous section, we use it to determine the structural failure condition of a given element. 
Here, we apply the Drucker-Prager yield criterion, a smooth approximation of the Mohr-Coulomb yield criterion \citep{Chen1988}:

\begin{equation}
    f = \alpha I_1 + \sqrt{J_2} - s \leq 0.
\end{equation}

\noindent $I_1$ and $J_2$ are the stress invariants:

\begin{equation}
    I_1 = \sigma_1 + \sigma_2 + \sigma_3,
\end{equation}

\begin{equation}
    J_2 = \frac{1}{6} \{(\sigma_1 - \sigma_2)^2 + (\sigma_2 - \sigma_3)^2 + (\sigma_3 - \sigma_1)^2\},
\end{equation}

\noindent where $\sigma_i$ ($i = 1, 2, 3$) is the principal stress component, which can be obtained from the
derived stress field above. $\alpha$ and $s$ are material constants and given by \citep{Chen1988}:

\begin{equation}
    \alpha = \frac{2 \sin{\phi}}{\sqrt{3} (3-\sin{\phi)}},
\end{equation}

\begin{equation}
    s = \frac{6Y \cos{\phi}}{\sqrt{3} (3-\sin{\phi)}}.
\end{equation}

\noindent From the equal condition of Equation (14), we obtain the following expression for $Y^*$:

\begin{equation}
    Y^* = \frac{\sqrt{3} (3 - \sin{\phi})}{6\cos{\phi}}\left(\alpha I_1 + \sqrt{J_2} \right).
\end{equation}

\noindent If Equation (19) becomes negative, we consider $Y^*$ to be $0 \ Pa$, which means that no
strength is necessary for an element to keep the original shape.

\section{Results} \label{sec:Results}

We investigate the critical cohesive strength based on the following assumed parameters, $\nu = 0.25$, $E = 10^7 \ Pa$,
and $\phi = 35 \degree$, to make our discussion consistent with earlier work (e.g., \citealt{Hirabayashi2015b}). We note
that Young’s modulus does not influence our stress field calculation \citep{love2013treatise}, and the variations in
Poisson’s ratio and the angle of friction do not affect our results for geological materials significantly
\citep{Lambe1969, Hirabayashi2019}. The current rotation period is set to be $3.60 \ hr$ \citep{Taylor2019}. We consider
the bulk density to be $0.5$, $1.0$, and $1.5 \ g/{cm}^3$. Table 1 lists the parameters considered in the current study.

We find that the failure mode varies with the bulk density. We plot the distribution of $Y^*$ at the $x$-$z$ plane for
different bulk densities in Figure 1. Panels a, b, and c describe the bulk densities of $0.5$, $1.0$, and 
$1.5 \ g/{cm}^3$, respectively. In this range of the bulk density, the body should have a cohesive strength to keep the
original shape. For the case of $\rho = 0.5 \ g/{cm}^3$, $P_c$ is found to be $4.83 \ hr$. Therefore, the current rotation
period is shorter than $P_c$, indicating that materials should be shed and highly sensitive to structural failure. Figure 1a 
shows $Y^*>0$ everywhere except the pole region. $Y^*$ is higher in the interior than on the surface, and its maximum
value is $259 \ Pa$ at the center. This indicates that the interior is more sensitive to structural failure than the
surface. If the actual cohesive strength $Y$ is smaller than $Y^*$, the central region structurally fails first. For the
case of $\rho = 1.0 \ g/{cm}^3$, $P_c$ is found to be $3.42 \ hr$ and is shorter than the current period. However, the
interior still exhibits high $Y^*$ in the major regions (Panel b). The maximum value of $Y^*$ is $179 \ Pa$ and is located
at the center. For the case of $\rho = 1.5 \ g/{cm}^3$, $P_c$ is found to be $2.79 \ hr$. Unlike the other two cases, the
interior has $Y^*=0$ in the most areas; however, high $Y^*$ is still distributed beneath the surface (${\sim}50 \ Pa$)(Panel c).

All these three cases show the sensitivity of Phaethon to structural failure. This body needs a cohesive strength to keep
the original condition without shape deformation. However, if there is a trigger of reshaping, it is likely that the deformation
process would be enhanced by rotation, as seen from the derived sensitivity. Because of the observed activities, the
cohesive strength of Phaethon is ${\sim}50 \ Pa \ - \ {\sim}260 \ Pa$, depending on the bulk density. This range is
consistent with that of observed small bodies \citep{Scheeres2018}.

\begin{deluxetable}{cCCC}[h!]
\tablenum{1}
\tablecaption{Parameter settings}
\tablewidth{0pt}
\tablehead{\colhead{Parameter} & \colhead{Symbol} & \colhead{Value} & \colhead{Units}}
\startdata
Gravitational constant  & G        & 6.6738 \times 10^{-11} & m^{3} \cdot kg^{-1} \cdot s^{-2}\\
Semi-major axis         & a        & 3200                   & m\\
Semi-minor axis         & c        & 2847                   & m\\
Oblateness              & \epsilon & 0.889                  & -\\
Current rotation period & P        & 3.60                   & hr\\
Bulk density            & \rho     & 0.5, 1.0, 1.5          & g \cdot cm^{-3}\\
Poisson's ratio         & \nu      & 0.25                   & -\\
Elastic modulus         & E        & 10^{7}                 & Pa\\
Friction angle          & \phi     & 35                     & deg
\enddata
\end{deluxetable}

\begin{figure}[h!]
\epsscale{0.7}
\plotone{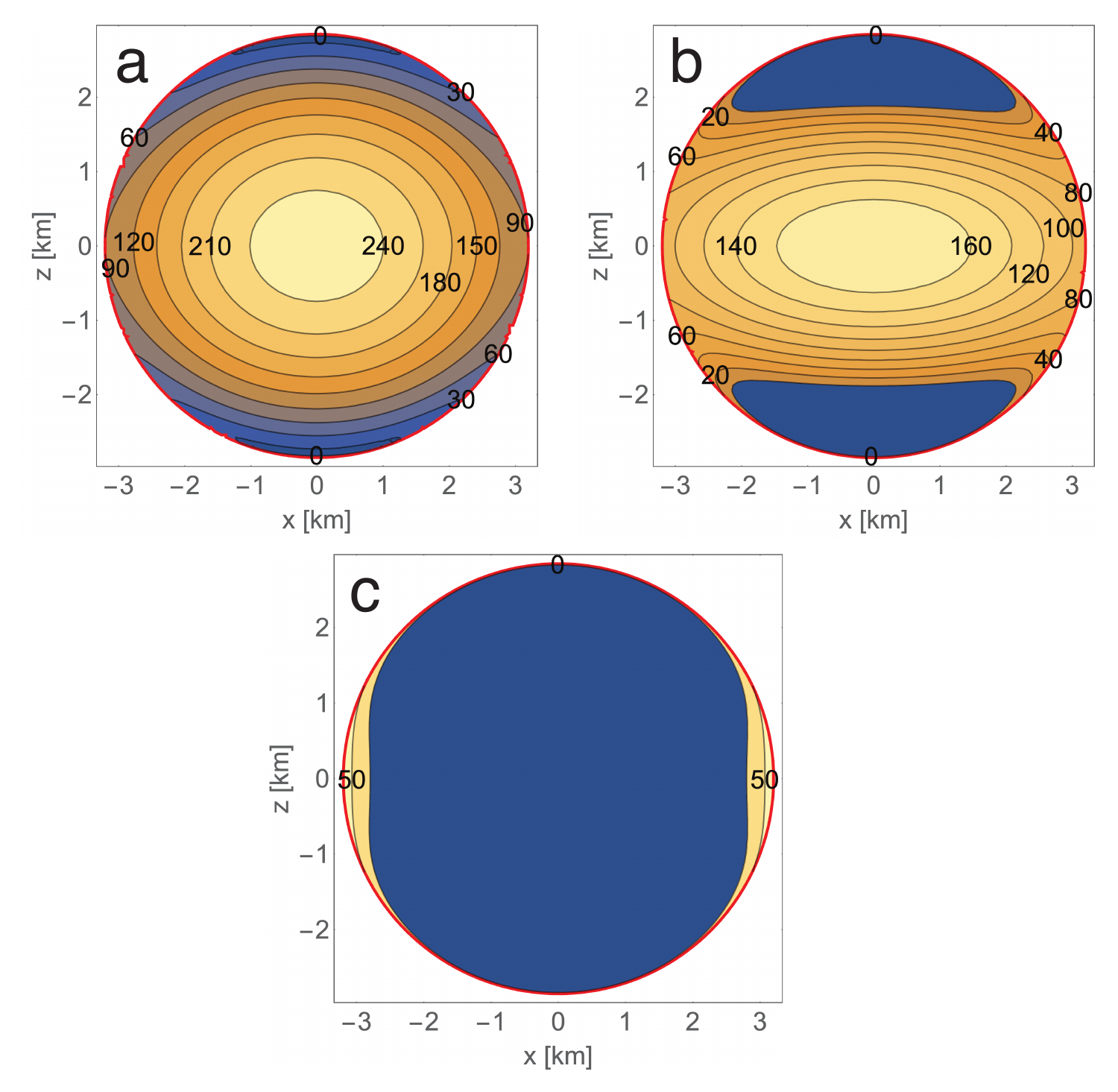}
\caption{Distribution of $Y^*$ at the $x$-$z$ plane. The rotation period is $3.60 \ hr$. Panels a, b, and c describe bulk densities of $0.5$, $1.0$, and $1.5 \ g/{cm}^3$, respectively.}
\end{figure}

\section{Discussions} \label{sec:discussions}
\noindent \textbf{\textit{Generation of dust tails at present}}

During perihelion passage in 2009 and 2012, the observations of Phaethon's dust tails revealed that this asteroid was an
active asteroid \citep{Jewitt2010, Li2013}. While the detailed mechanisms are not well known, thermal waves may be one of
plausible explanations for the reported mass shedding activities \citep{Li2013}.

At the current rotation period of $3.60 \ hr$, the body is sensitive to structural failure regardless of the bulk density and thus
needs cohesive strength to maintain its shape. The derived cohesive strength of Phaethon is less than 
${\sim}50 \ Pa \ - \ {\sim}260 \ Pa$, which is consistent with that of small bodies ranging up to a few hundred pascals
\citep{Scheeres2018}. We interpret this sensitivity as a potential enhancement of mass shedding. If there is a trigger of
reshaping even at small scales, the structure of Phaeton would be perturbed, leading to rotationally driven reshaping at
larger scales. Such a trigger can be thermal waves in thin surface layers \citep{Jewitt2010, Jewitt2012, Li2013}.
Micrometeoroid impacts or other processes may also be possible, as seen on Bennu \citep{Lauretta2019}, although thermal
waves, again, are more consistent to explain the activities of Phaethon around its perihelion passage \citep{Li2013}.\\

\noindent \textbf{\textit{Possible source of the Geminid meteor stream}}

We expect that rotationally induced structural failure makes Phaethon more oblate, i.e., $\epsilon$ becomes lower
\citep{Hirabayashi2015}. Because the angular momentum is constant during deformation, Phaethon may have been less oblate
and rotated faster at an earlier stage before it had large deformation. Figure 2 shows the dependence of $P$ on $\epsilon$
and $P_c$ for different bulk densities. We find that if Phaethon is less oblate, the rotation period is shorter, and thus
$Y^*$ should be higher at a shorter rotation period. If Phaethon is a sphere ($\epsilon =1.0$; $a = b = c$), which is the
shape condition that is the least affected by rotation, the rotation period should become $P = 3.38 \ hr$. Figure 3 shows
the distribution of $Y^*$ for this case. Similar features observed in Figure 1 are evident in this case; however, $Y^*$ is
higher, implying that Phaethon should have been more sensitive to structural failure at a shorter rotation period. 
Therefore, the failure mode may become severer, and more materials should be shed
significantly at a shorter rotation period.

\begin{figure}[ht!]
\epsscale{0.8}
\plotone{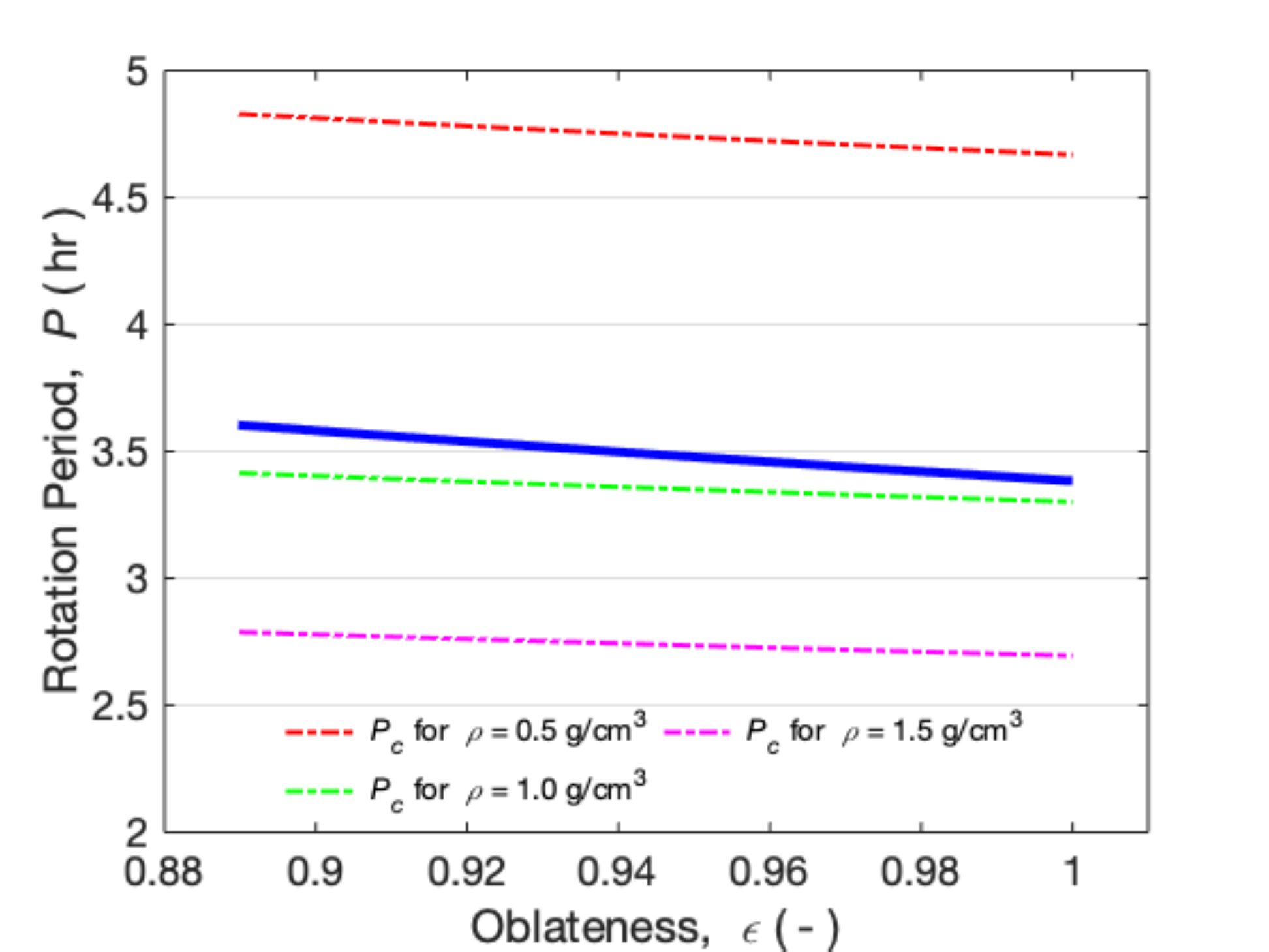}
\caption{Rotation period as a function of the oblateness (the blue solid line). $P = 3.60 \ hr$ when $\epsilon = 0.889$
and $P = 3.38 \ hr$ when $\epsilon = 1.0$. The dotted lines indicate $P_c$ for different bulk densities: the red, green, and magenta show bulk densities of $0.5$, $1.0$, and $1.5 \ g/{cm}^3$, respectively.}
\end{figure}

\pagebreak 

\begin{figure}[ht!]
\epsscale{0.7}
\plotone{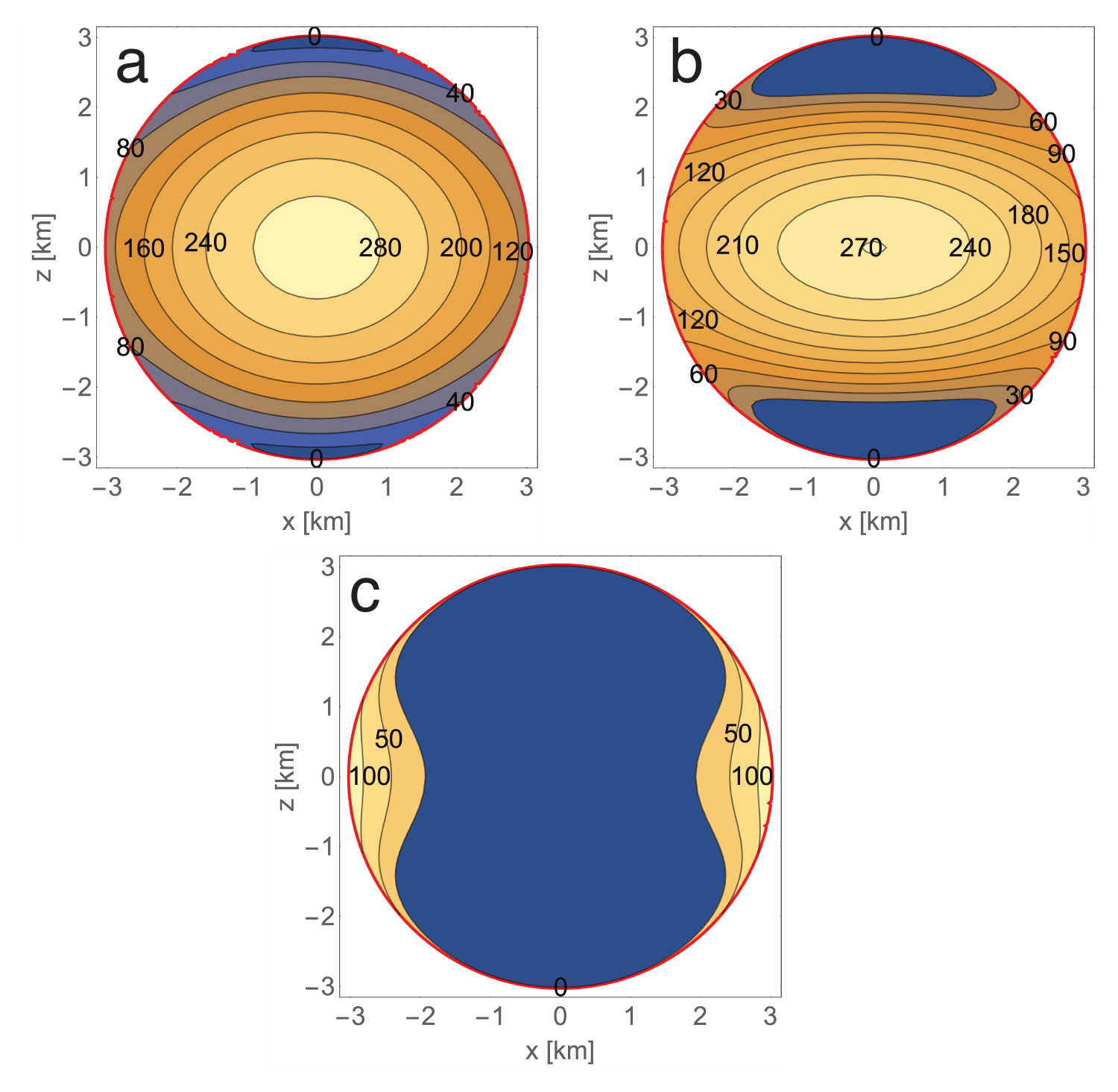}
\caption{Distribution of $Y^*$ at the $x$-$z$ plane. The rotation period is $3.38 \ hr$. Panels a, b, and c describe bulk densities of $0.5$, $1.0$, and $1.5 \ g/{cm}^3$, respectively.}
\end{figure}

From these results, we propose a possible evolution scenario of Phaethon (Figure 4).
Phaethon was originally less oblate and spinning at a shorter rotation period than the current period.
This stage is before the Geminid meteor stream was generated.
Possible initiation processes such as thermal waves during Phaethon’s apparition passages triggered reshaping,
and rotational deformation enhanced this reshaping process significantly.
Because the rotation period at this stage was closer to or above $P_c$,
the reshaping process caused mass shedding at large scale, which became a source of the Geminid meteor stream.
Thus, the current oblate shape maybe a remnant of this large deformation event.
When the oblateness evolved, the rotation of Phaethon slowed down.
However, the structure was still sensitive to failure. 
When there is similar perturbation such as thermal waves at present,
rotationally driven failure can be triggered; 
however, because the centrifugal effect is less significant,
the magnitude of mass shedding is less intense than that in the past.

While we cannot strongly constrain whether the generation of the Geminid meteor stream was a single event or episodic, 
our study gives some hints of large-scaled mass shedding processes as a source of the Geminid meteor stream.
Because the generation of the Geminid meteor stream may have occurred within the last $1 \ ka$ 
\citep{Gustafson1989, Williams1993, Ryabova2007}, the large-scaled reshaping and mass shedding processes
may have occurred in this timescale. 
These processes should be more rapid and intense to be completed (within the last $1 \ ka$) 
than the YORP driven evolution, which may take $ {\sim}1 \ Ma$ based on earlier work \citep{Capek2004, Bottke2006}.
Given this short timescale, a possible explanation of the deformation mode may be internal failure,
which can provide large-scaled deformation and thus mass shedding at fast rotation \citep{Hirabayashi2015}.

We note that the total mass of the Geminid meteor stream is about $2.5\%$ of that of Phaethon. 
If a mass shedding even that can produce the same magnitude of the Geminid meteor stream occurs every $1 ka$, 
the lifetime of Phaethon would only be ${\sim}40 \ ka$,
which may be much shorter than that predicted by the orbital evolution, ${\sim}26 \ Ma$ \citep{DeLeon2010}.
While there is no decisive evidence, the YORP effect may give some insights into this discrepancy.
The YORP evolution timescale of Phaethon may become longer by many different factors
such as the mass, shape, surface composition, and stochastic evolution \citep{Bottke2015}.
Furthermore, its highly eccentric orbit, $e = 0.890$, with small perihelion distance of 0.14 AU
(JPL Small-Body Database) may give strong variations in solar radiation acting on Phaethon.
Thus, it may be possible that Phaethon has stochastically spun up by the YORP effect for the entire orbital age,
${\sim}26 Ma$ \citep{DeLeon2010}, and it recently experienced large-scaled mass shedding
that formed the Geminid meteor stream.
Then, a large mass shedding event that produced the Geminid meteor stream may have decelerated Phaethon's spin,
but the spin state after this event may have been still high enough
to have some mass shedding events at small level, similar to what we observed in 2009 and 2012
\citep{Jewitt2010, Jewitt2012, Li2013}.
To fully address the detailed timescale of the rotationally driven reshaping process
is beyond the scope of this study; we leave this problem as future work.

Finally, We assumed that Phaethon currently has a top shape having Bennu’s oblateness,
$\epsilon = 0.889$, by following \citet{Taylor2019}.
To check if this setting is consistent with other top-shaped asteroids, we consider six top-shaped
near-earth asteroids: Ryugu \citep{Watanabe2019}, 1994 KW4 \citep{Ostro2006}, 2008 EV5 \citep{Busch2011}, 1994 CC
\citep{Brozovic2011}, 2001 SN263 \citep{Becker2015}, and 2000 DP107 \citep{Naidu2015}. 
We find the range of $\epsilon$ to be $0.873$ – $0.968$, and thus Bennu’s $\epsilon$ is within this range.
We conduct the same analyses for these objects and found trends similar to Bennu’s.
The variation in the oblateness at this magnitude does not affect the results of $Y^*$ significantly,
which is less than 12\%. 
Therefore, we conclude that our oblateness setting is meaningful to capture a possible scenario of Phaethon’s activities.

\begin{figure}[ht!]
\epsscale{0.8}
\plotone{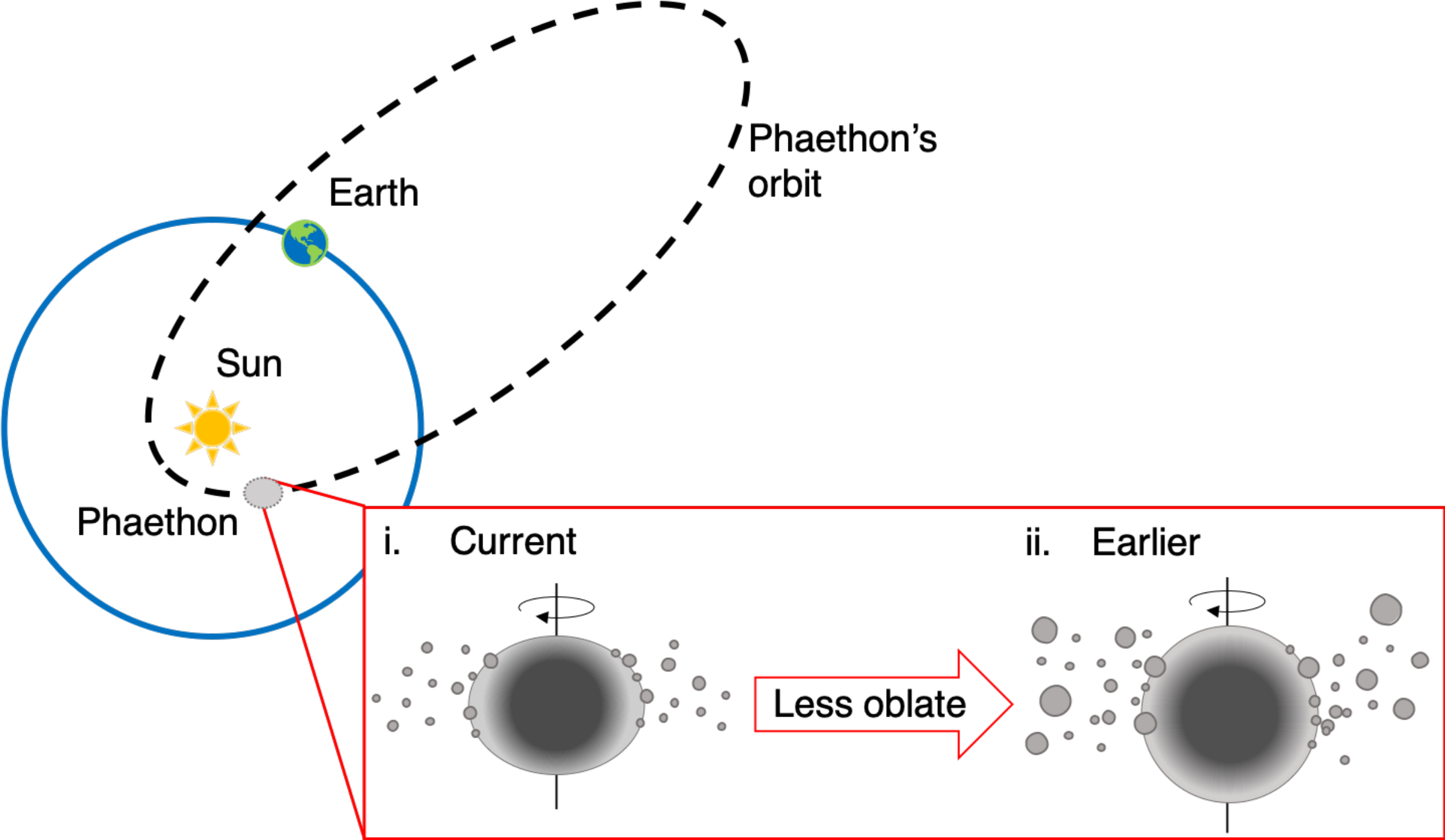}
\caption{Possible evolution scenario of Phaethon. Phaethon has a highly eccentric orbit ($e = 0.890$) with small
perihelion distance of 0.14 AU. At an earlier stage, Phaethon was less oblate and rotating faster, leading to mass
shedding at larger scale. For both stages, perturbation such as thermal waves may be a possible trigger for mass shedding.}
\end{figure}

\noindent \textbf{\textit{Potential issues}}

We finally address issues of our analysis approach. This study explored Phaethon’s rotationally induced structural failure
by modeling Phaethon as a triaxial ellipsoid. We did not account for local topographic features; therefore, our
semi-analytical model does not capture local deformations which may differ from global deformations
\citep{Hirabayashi2019, hirabayashi2019western}. However, \citet{Hirabayashi2015}, who compared an analytical solution and a FEM solution, found
that there was no significant variation between the two. Therefore, we conclude that our semi-analytical model can provide
solutions with reasonable accuracy. It is our future work to perform FEM analysis using a shape model to investigate the
local deformations.

Another issue to be addressed is that our model does not assert the detailed failure mode. While numerous research studies
have been undertaken, it is still not well determined how top shape asteroids deform. \citet{Hirabayashi2015} showed that
depending on the internal structure and the bulk density, the failure mode may become different – either surface
processing or internal deformation. However, because our semi-analytical model can only describe a homogeneous structure,
we cannot infer how Phaethon's deformation is controlled by the heterogeneity. Therefore, the detailed failure mode cannot
be specified. If Phaethon’s heterogeneous structure is revealed, we need to employ different approach (i.e.
\citealt{Hirabayashi2015a}).

There still remain many unknowns regarding Phaethon's physical properties. We will further elaborate our approach to give
constraints on the activity of Phaethon. Concurrently, further constraints are vitally important to support DESTINY+.

\acknowledgments
RN and MH acknowledge support from NASA/Solar System Workings (NNH17ZDA001N/80NSSC19K0548) and Auburn
University/Intramural Grant Program.


\end{document}